\documentclass[aps,12pt,tightenlines,amsmath,amssymb,showpacs]{revtex4}
\usepackage{graphicx}
\usepackage{dcolumn}
\usepackage{bm}
\usepackage{epsfig}

\newcommand{\labeld}[1]{ }

\newcounter{remark}
\newcommand{\be}{\begin{equation}}\newcommand{\ee}{\end{equation}}
\newcommand{\bea}{\begin{eqnarray}}\newcommand{\eea}{\end{eqnarray}}
\def\Eqref#1{Eq.~(\ref{#1})}

\def\newremark#1{\refstepcounter{remark} \label{r.#1}\smallskip\noindent\textsf{Remark \arabic{remark}}:~\labeld{r.#1}}   

\def\sgn{\mathop{\rm sgn}\nolimits}

\def\Dx{\partial_x}

\def\Dy{\partial_y}
\def\Dz{\partial_z}

\def\Dt{\partial_t}

\def\M{{\cal M}}
\def\M{\hbox{${\cal M}$}}
\def\Mz{\hbox{${\cal M}_z$}}
\def\Mzmath{{\cal M}_z}

\def\DDrp{{\bm \nabla_{\rr'}}}
\def\DDr{{\bm \nabla_{\rr}}}
\def\DD{{\bm \nabla}}

\def\kappab{{\bm \kappa}}
\def\kk{{\bm k}}
\def\pdot{{\bm\dot\pp}}
\def\pp{{\bm p}}
\def\PP{{\bm P}}
\def\qq{{\bm q}}
\def\rhob{{\bm \rho}}
\def\RR{{\bm R}}
\def\rr{{\bm r}}

\def\Rbb{{\mathbb R}}

\def\ex{{\bm {\hat e}_1}}
\def\ey{{\bm {\hat e}_2}}
\def\ez{{\bm {\hat e}_3}}

\def\Ltwoplusthree{\hbox{L$^2_+(\Rbb^3)$}}
\def\Ltwominusthree{\hbox{L$^2_-(\Rbb^3)$}}
\def\Ltwoplusminusthree{\hbox{L$^2_\pm(\Rbb^3)$}}
\def\Ltwo{\hbox{L$^2$}}
\def\Ltwothree{\hbox{L$^2(\Rbb^3)$}}

\def\LtwoMzprime{\hbox{L$^2({\cal M}_z')$}}

\def\LtwoMz{\hbox{L$^2(\M_z)$}}

\def\Sone{\textsf{S}$^1$}

\def\qdot{{\bm \dot \qq}}

\def\intline{\int_{-\infty}^\infty}
\def\intplane{\int_{\Rbb^2}}

\def\noteoperational{If one is desperate to restore the conventional Hilbert space paradigm, there is an approach to arrival times in which one need not define a time operator. In practice one can measure time by measuring positions \cite{detectormodelmovingparticles, peres, wigner}. In this way the ``operator'' for time becomes the collection of operators for the position measurements. A similar viewpoint can be adopted for the momentum problem}

\def\notemultiple{An example of path that is not deformable to a point is one that goes from $(1,1,1)$ to $(-1,-1,-1)$\@. When it (necessarily) hits the $z=0$ plane it jumps to its parity reflection. Since $(0,0,0)$ is excluded, this jump cannot be eliminated. The space \M\ is also not orientable}

\begin{document}
\title{Relative momentum for identical particles}

\author{B. Gaveau}
\affiliation{Laboratoire analyse et physique math\'ematique, 14 avenue F\'elix Faure, 75015 Paris, France}
\author{L. S. Schulman}
\affiliation{Physics Department, Clarkson University, Potsdam, New York 13699-5820, USA}
\email{schulman@clarkson.edu}

\date{\today}
\begin{abstract}
Possible definitions for the relative momentum of identical particles are considered.
\end{abstract}

\maketitle

The mantra of quantum mechanics is that an observable is a self-adjoint operator and its eigenvalues are the possible results of experiments. This criterion is neither necessary nor sufficient. That it is not sufficient is manifest in the fourth chapter of Gott\-fried's 1966 book \cite{gottfried} where he points out that perfectly good operators for what are now called Schr\"odinger cats or grotesque states \cite{griffiths} are not observable. That it is not necessary one learns from that most precisely measured of physical quantities, time, which resists definition as a self-adjoint operator~\cite{muga, note:operational}.

In this article we find that another meaningful physical quantity, the relative momentum of identical particles, represents a further failure of the general framework. Momentum is already known to be problematic in two cases: the (single) hard wall \cite{bonneau} and radial momentum~\cite{messiah}. However, an infinite wall is an idealization and the difficulties in the second case could be attributed to the choice of coordinates. With Cartesian coordinates there is no problem.

But the relative momentum of a pair of identical particles is unavoidably fundamental. The concept is not straightforward either physically or mathematically. If you cannot know which is which (more precisely, it is simply not defined), how can you attribute a vector to the difference. Mathematically, in the conventional way of dealing with identical particles one assigns identities (say \#1 and \#2), but works either on the space of symmetric or skew-symmetric states. However, the operator $\pp_1-\pp_2$ (relative momentum) applied to a state takes you from one space to the other, and is thus not defined in the relevant Hilbert space. On the other hand, there is nothing wrong with the \textit{square} of the relative momentum, nor with the squares in the examples of the previous paragraph.

Physically, however, $(\pp_1-\pp_2)^2$ is not enough. Consider an experiment on a pair of electrons. One can measure momentum with two pairs of position detectors, A and B\@. They are located so that it is overwhelmingly likely that energy constraints imply that the deduced $\pp_A$ and $\pp_B$ correspond to the individual electrons. The center of mass momentum, $\PP\equiv \pp_A+\pp_B$ is a symmetric well-defined operator, which conventionally is called $\pp_1+\pp_2$, giving the electrons identities that carry no significance. However, equivalence under exchange means that neither $(\pp_1-\pp_2)/2$ nor $(\pp_2-\pp_1)/2$ can be the measured $\pp_{_{AB}} \equiv(\pp_A-\pp_B)/2$\@. The equivalence suggest that $(\pp_1-\pp_2)/2$ and $(\pp_2-\pp_1)/2$ can be represented by the same point on the projective (2-) sphere (a sphere with antipodal points identified). But even this is less than the measurement has revealed, which includes information on whether the electrons are approaching or receding. There should thus be a concept of time-derivative of this momentum.

In this article we show how the projective characterization arises naturally from a systematic identity-blind treatment of identical particles. This will also allow us to define Hilbert space objects associated with relative momentum, although these objects do not conform to the paradigm connecting self-adjoint operators and observables. We emphasize that the measurements of  $\pp_A$ and $\pp_B$, and functions thereof, are perfectly well-defined. It is the theory that is inadequate.

We take two approaches. The first has the conceptual advantage that the very language precludes distinguishing the particles. Moreover, we do not need to define any new operators: the unitary propagator is known, and its logarithm gives us the effective Hamiltonian. (For the conceptual issues we confront, the particles are taken to be non-interacting.) The second works within the usual framework of even or odd functions, and we find an appropriate operator for relative momentum. It is not self-adjoint and thus not a conventional observable. In general it is not even Hermitian. But we show that it is the physically correct object. The two developments are mutually consistent.

Terminology: We call $A$ \textit{Hermitian} if $\langle \phi|A|\psi\rangle=\langle \psi|A|\phi\rangle^*$ with $\phi,\psi\in D_A=\,$the domain of $A$\@. The term ``symmetric'' (sometimes used for this concept) will be restricted to exchange properties of functions. The word ``identity'' is used in two senses: the name of the particle and the notion that two ``identical'' particles can in principle not be distinguished, hence have no identity.

\textit{Changes in momentum: using the propagator.~~}
Two identical particles in 3-space can be described by a pair $(\rr_1,\rr_2)$ ($\rr_k\in\Rbb^3$, $k=1,2$). But this description is restricted: first, conceptually, $(\rr_1,\rr_2)$ and $(\rr_2,\rr_1)$ must be considered the same point; second, if $\rr_1=\rr_2$ the pair is not a pair and the object is meaning\-less. The appropriate space can thus be written $\left\{(\rr_1,\rr_2)\in\Rbb^6\,|\; \rr_1\ne\rr_2 \right\}/\left[(\rr_1,\rr_2)\sim(\rr_2,\rr_1)\right]$\@. In words, this is 6-dimensional Euclidian space, minus the ``diagonal'' ($\rr_1=\rr_2$), modulo equivalence under exchange. By a change of coordinates this space is $\widetilde\M \equiv\Rbb^3\times\M$ with $\M\equiv \{\Rbb^3-\{{\bm0}\}\}/[\rr\sim-\rr]$\@. As for other such spaces (e.g., the Klein bottle, the rotation group) it is convenient to have a representation in ordinary space. We choose
$\widetilde\M_z \equiv\Rbb^3\times\M_z$, with
\be
\M_z=\{\rr=(x,y,z)\in\Rbb^3\,|\, (z>0) \vee (z=0, y>0) \vee (z=y=0, x>0) \}
\,.
\label{e.Mdefinition}
\ee
\labeld{e.Mdefinition}
The subscript $z$ on \M\ indicates this choice of coordinates. Connecting this to the original description, the first space in the product is the center of mass coordinate, $\RR\equiv(\rr_1+\rr_2)/2$, and the second (\M\ or $\M_z$) the space of relative coordinates, $\rr\equiv\rr_1-\rr_2$\@. We focus on \M\@.

\smallskip

\newremark{pisdifferent}
Although coordinate space is \M, momentum space is not. Besides the information that \M\ carries there is the issue of whether the particles are approaching or receding. Also, as we will see, there is no self-adjoint operator whose spectrum would be momentum space.

\newremark{spin}
We focus on coordinate space wave functions, assuming that the spin state of the particles is either symmetric or skew-symmetric and is factored out. Thus our symmetric or skew-symmetric wave functions can represent either bosons or fermions.

\smallskip

The spaces $\widetilde\M$ and \M\ are multiply connected \cite{note:multiple}, creating an ambiguity in the path integral representation of the propagator. The way to deal with this was discussed in \cite{thesis, pispin, approximate, laidlaw, lsspibookallshort} and involves, for scalar states, commutative representations of the fundamental homotopy group. In this case the group is $Z_2$ \cite{laidlaw}\@. The prescription of Ref.\ \cite{thesis} is to go to the covering space of \M, take each preimage of the initial point, and evaluate its propagator to the final point (using the dynamics induced by the inverse of the covering projection, assuming requisite smoothness). Then these propagators are added, with phases determined by the representations of the group. Therefore, on $\widetilde\M$, depending on the symmetry of the spin state and the nature of the identical particle (fermion or boson), the propagator is
\be
G(\rr_1'',\rr_2'',t;\rr_1',\rr_2')=G_0(\rr_1'',\rr_2'',t;\rr_1',\rr_2')
                   \pm G_0(\rr_1'',\rr_2'',t;\rr_2',\rr_1')
\label{e.freepair}
\,.
\ee
\labeld{e.freepair}
$G_0$ is the propagator on the covering space (so the second term in \Eqref{e.freepair} is well-defined). Because $G$ is used only in integrals we extend its argument to the entire $z=0$ plane, since this only adds sets of measure zero. Finally, functions smooth enough to have a $z\to0$ limit necessarily have the same symmetry as the propagator in the plane $z=0$\@.

For non-interacting particles
\be
G_0(\rr_1'',\rr_2'',t;\rr_1',\rr_2')=g_m(\rr_1''-\rr_1',t)\,g_m(\rr_2''-\rr_2',t)
\label{e.G0ident}
\ee
\labeld{e.G0ident}
with $g_m$ defined by ($\hbar=1$, $m=$ particle mass)
\be
g_m(\rr,t)\equiv ( m/(2\pi it))^{3/2}\;e^{im|\rr|^2/2t}
\label{e.gvec}
\,.
\ee
\labeld{e.gvec}
Defining $M=2m$ and $\mu=m/2$, it follows from \Eqref{e.gvec} that
\be
G(\rr_1'',\rr_2'',t;\rr_1',\rr_2')=
        g_M(\RR''-\RR',t)
                 \left[g_\mu(\rr''-\rr',t)\pm g_\mu(\rr''+\rr',t)  \right]
\,.
\label{e.GCM}
\ee
\labeld{e.GCM}
On \Mz, $\pp$ takes its usual form; the changes in the propagator modify the \textit{dynamics}. We focus on \Mz, involving only the relative coordinate. To evaluate changes in $\pp$ we calculate (taking $\mu=1$ and suppressing the subscript on $g$)
\bea
([\pp,U]\psi)(\rr)&=&\intline\!\!\! dx'\! \intline\!\!\! dy'\! \int_{z'>0}\!\! dz'
              \biggl[\frac1i\DDr \left[g(\rr-\rr',t)
              \pm
              g(\rr+\rr',t)\right]\psi(\rr')  \nonumber\\
                  &&\qquad\qquad    -  \left[g(\rr-\rr',t)
                  \pm
                  g(\rr+\rr',t)\right]\frac1i\DDrp\psi(\rr')  \biggr]
\,,
\label{e.pUcommutator3}
\eea
\labeld{e.pUcommutator3}
where $U$ is the unitary operator whose kernel is $G$\@. We next switch $\DDr$ to $\DDrp$ and perform a number of integrations by parts. We also make use of the vanishing of integrals whose overall symmetry is odd. This yields
\bea
([\pp,U]\psi)(\rr)&=&-i \ez
\left.\intline\!\!\! dx'\! \intline\!\!\! dy'
              \left[ g(\rr-\rhob',t)\mp g(\rr+\rhob',t)\right]\psi(\rhob')\right|_{z=0}
                     \nonumber\\
         &&\qquad  \pm    2i\intline\!\!\! dx'\! \intline\!\!\! dy'\! \int_{z'>0}\!\! dz'
                  g(\rr+\rr',t)\DDrp\psi(\rr')
\,,
\eea
where $\rhob\equiv x\ex+y\ey$ and $(\ex,\ey,\ez)$ are unit vectors in the $(x,y,z)$ directions. The first integral above vanishes because for both wave function symmetries the combination of $g$'s and the wave function have opposite parity in the plane.

Next take the inner product with a function, $\phi$, having the same symmetry properties as $\psi$, and consider the small time limit:
\bea
\langle\phi|[\pp,U]|\psi\rangle &=&
           \pm 2\cdot2i
         \intline\!\!\! dx'\! \intline\!\!\! dy'\! \int_{z'>0}\!\! dz'
         \intline\!\!\! dx''\! \intline\!\!\! dy''\! \int_{z''>0}\!\! dz''
                  \nonumber\\
       &&\qquad\qquad        \phi^*(\rr'')   g(\rr''+\rr',t)\DDrp\psi(\rr')
\,.
\eea
The additional 2 above arises because $\langle\phi\,|\psi\rangle=2\int\phi^*\psi$, a property demanded by the correspondence with the usual representation (see below). The integral over $\rhob''$ can be performed using $g(\rhob'+\rhob'',t)\sim\delta(\rhob'+\rhob'')$, for small enough time. Using the symmetry or skew-symmetry of $\phi$ one obtains
\bea
\langle\phi|[\pp,U]|\psi\rangle &=&
          4i
         \intline\!\!\! dx'\! \intline\!\!\! dy'\! \int_{z'>0}\!\! dz'
          \int_{z''>0}\!\! dz''
                  \nonumber\\
       &&\qquad\qquad        \phi^*(\rhob'-z''\ez)   g(z''+z',t)\DDrp\psi(\rr')
\,.
\eea
Since $g(z''+z',t)$ acts like a $\delta$-function in $z$ (for small $t$), we expand $\phi$ and $\psi$ around $z=0$ to get
\bea
\langle\phi|[\pp,U]|\psi\rangle &=&
           4i
         \intplane\!\!\! d^2\rhob'\!  \int_{z'>0}\!\! dz'
          \int_{z''>0}\!\! dz''
             g(z''+z',t)      \nonumber\\
       &&\qquad  \times \left[     \phi^*(\rhob') -z''\Dz\phi^*(\rhob')+\frac{z''^2}2 \Dz^2 \phi^*(\rhob'')+\dots\right]
       \nonumber\\
       &&\qquad\quad\times
             \left[ \DDrp\psi(\rhob')+z'\Dz\DDrp\psi(\rhob')+\frac{z'^2}2\Dz^2\DDrp\psi(\rhob')+\dots  \right]
\label{e.expand}
\,.
\eea
\labeld{e.expand}
In \Eqref{e.expand} the zeroth and second order terms in $z$ drop out because they involve products of functions of opposite parity. The remaining first order terms in $z'$ and $z''$ are
\bea
\langle\phi|[\pp,U]|\psi\rangle &=&
           4i
         \intplane\!\!\! d^2\rhob'\!  \int_{z'>0}\!\! dz'
          \int_{z''>0}\!\! dz''
           g(z''+z',t)      \nonumber\\
       &&\qquad  \times  \left \{  \phi^*(\rhob')
        z'\Dz\DDrp\psi(\rhob') - z''\Dz\phi^*(\rhob') \DDrp\psi(\rhob')  \right\}
\,.
\eea
To evaluate this we require the integral
$f(t)\equiv (2\pi it)^{-1/2} \int_0^\infty du\, u\int_0^\infty dv\, e^{i(u+v)^2/2t}$\@.
By standard manipulations one finds that $f(t)=it/4$ for $t\to0$\@. Therefore
\be
\langle\phi|[\pp,U]|\psi\rangle
        =
         - t
         \intplane\!\!\! d^2\rhob'\!
                \left \{     \phi^*(\rhob')
        \Dz\DDrp\psi(\rhob') -\Dz\phi^*(\rhob') \DDrp\psi(\rhob')  \right\}
\,.
\ee
Since $U$ is unitary, for small $t$ $U(t)=1-iH_\mathrm{eff}t$. Then
\bea
\langle\phi|\pdot|\psi\rangle &\equiv& \Dt \langle\phi|\pp|\psi\rangle
               =-i\langle\phi|[\pp,H_\mathrm{eff}]|\psi\rangle \nonumber\\
        &=&    -
                \intplane\!\!\! d^2\rhob'\!
                \left \{     \phi^*(\rhob')
                \Dz\DDrp\psi(\rhob') -\Dz\phi^*(\rhob') \DDrp\psi(\rhob')  \right\}
\,.
\label{e.pdotmatrixelement31}
\eea
\labeld{e.pdotmatrixelement31}

\newremark{terrible}
$\pdot$ is not a densely defined Hilbert space operator. Its action on a function $\psi$ would necessarily multiply $\psi$ by $\delta(z)$ or even $\delta'(z)$.

\textit{Conventional wave functions, unconventional momentum.~~}
The usual treatment of two identical particles uses symmetric or skew-symmetric wave functions on $\Rbb^6$\@. Eliminating center of mass coordinates, the Hilbert space is L$^2_\pm(\Rbb^3)$\@. Relative momentum (or any relative coordinate), as an operator, maps one \textit{out of} the appropriate space. For example, if $\psi$ is odd, $p_z\psi$ is even. (Alternatively, $\pp$ has formal matrix element zero between any two states.) However, on L$^2(\Mz)$ one does not have this problem. We now show how an isometry between the Hilbert spaces gives a natural candidate for relative momentum on L$^2_\pm(\Rbb^3)$\@.

On L$^2(\Mz)$ we use the scalar product $\langle\phi_1|\phi_2\rangle = 2\int_{\Mzmath}\phi_1(\rr)^* \phi_2(\rr)\,d^3\rr $, since the wave functions are normalized to unity on $\Rbb^3$\@. Then we have isometries of the Hilbert spaces
\bea
V_\pm :&& \Ltwoplusminusthree \to \hbox{L}^2(\Mz) \nonumber\\
(V_\pm\psi)(\rr)&=&\psi(\rr)\,,\ \rr\in\Mz
\,,
\eea
with inverses given by
\be
(V_\pm^{-1}\phi)(\rr)=
             \begin{cases} \phi(\rr)  & \text{if $z>0$} \\
                          \pm \phi(-\rr) & \text{if $z<0$}
             \end{cases}
\,.
\ee
If $A\mskip 1mu$: $D_A\to \LtwoMz$ is a densely defined operator with $D_A\subset \LtwoMz$, we associate $A'_\pm\mskip1mu$: $D'_{A\pm}\to \Ltwothree$ by the formula $A'_\pm = V_\pm^{-1} A V_\pm$, with $D'_{A\pm}=V_\pm^{-1}(D_A)$\@. Take $A=-i(\Dx,\Dy,\Dz)$\@. Then for both kinds of statistics the corresponding operator is
\be
A'_\pm =-i\sgn(z)(\Dx,\Dy,\Dz)
\,.
\ee
The proof is an immediate consequence of the definitions.

 Let $\qq\equiv-i\sgn(z)\DD$\@. There are several implications of this result.

On \Ltwominusthree, $\qq_z$ is Hermitian (do an integration by parts) but not self-adjoint. On \Ltwominusthree\ the maximal domain on which $\qq_z$ is Hermitian is
$
D_-\allowbreak=\{f\in\hbox{\Ltwominusthree}\,\allowbreak|\,\allowbreak \Dz f\in\hbox{L}^2,\allowbreak f((x,y,0))=0 \}
$\@.
But its adjoint has domain
$
D^*_-\allowbreak=\left\{
f\in\hbox{\Ltwominusthree}\,\allowbreak\Big|\,\allowbreak\Dz (f|_{{\Mzmath}} )   \in\LtwoMz \,,\allowbreak\,
                                           \Dz (f|  _{{\Mzmath}'} )   \in\LtwoMzprime
                                           \right\}
$,
where $\Mz'$ is the complement of $\Mz$. Thus $\Dz f$ may not be in $\Ltwo$, since a (permissible) discontinuity at $z=0$ makes its derivative a distribution.

On \Ltwoplusthree, $\qq_z$ is not even Hermitian. An integration by parts of $\int_{-\infty}^{\infty}\phi^*\sgn(z)(-i)\Dz\psi$ gives $\int_{-\infty}^{\infty}(\Dz(-i)\phi^*)\sgn(z)\Dz\psi+2i\phi^*(0)\psi(0)$. Unlike the odd case, the additional contribution need not vanish. Taking a domain in which this (as well as derivatives at zero) vanish would lead to even more trouble, since for any finite energy the time independent Schr\"odinger equation (being second order) would force the function to vanish everywhere. On \Ltwoplusthree\ a natural domain for $\qq_z$ would be
$ D_+=\left\{\psi\in\Ltwoplusthree \,\big|\Dz\psi\in\Ltwo\right\} $\@.
But $-i\sgn(z)\Dz$ is not even Hermitian. Indeed the domain of a putative adjoint would be
$ D_+^* =\left\{\psi\in\Ltwoplusthree \,\big|\Dz\psi\in\Ltwo \,, \psi|_{z=0}=0  \right\} $\@.
This can be seen directly; alternatively if one formally symmetrizes $-i\sgn(z)\Dz$ you pick up a delta function in the integration by parts. Thus for formal symmetry, $\psi$ must vanish at $z=0$\@.

The form developed for $\qq$ allows a direct calculation of $\qdot$\@. On \Ltwoplusminusthree\ $H$ is simply $p^2/2$, unchanged from its free form, and it is the momentum that has changed, becoming $\qq\equiv-i\sgn(z)\DD$\@. We obtain $\qdot$ through the evaluation of $[H,\qq]$:
\be
[H,\qq]=\frac i2 [\Dx^2+\Dy^2+\Dz^2,\sgn(z) \DD]
=\frac i2 [\Dz^2,\sgn(z)]\DD
\label{e.enum50}
\,.\ee
\labeld{e.enum50}
Using $[\Dz,\sgn(z)]=2\delta(z)$ we find
$[\Dz^2,\sgn(z)] = 2\delta'(z)+4\delta(z)\Dz $\@. It follows that
\bea
\langle\phi|\qdot|\psi\rangle&=&i\langle\phi|[H,\qq]|\psi\rangle
\nonumber\\
&=&    \int_{\Rbb^2} d^2 \rhob \;\Dz \left(\phi^*(\rhob)    \DD \psi(\rhob) \right)
-2 \int_{\Rbb^3} d^3 \rr \phi(\rr)^* \delta(z)\Dz   \DD \psi(\rr)
\nonumber\\
&=&    \int_{\Rbb^2} d^2 \rhob \,\left[\Dz \left(\phi^*(\rhob) \right)\DD\psi(\rhob)
           - \phi^*(\rhob) \Dz  \DD \psi(\rhob)\right]
\label{e.enum53}
\,,
\eea
\labeld{e.enum53}
in agreement with \Eqref{e.pdotmatrixelement31}.

\textit{Verification.~~}
Let $\psi(\rr,t)\in\,$\Ltwoplusminusthree\ evolve under the free Hamiltonian. Then
\be
\psi(\rr,t)=\int \frac{d^3k}{(2\pi)^3}\, a(\kk)e^{i\kk\rr}e^{-i\kk^2t/2}
\ee
with $a(-\kk)=\pm a(\kk)$, according to the symmetry of $\psi$\@. For this $\psi$ (partially suppressing $t$) \Eqref{e.enum53} implies
\bea
\langle\psi|\qdot|\psi\rangle
&=&
\int \frac{d^3k\, d^3k'}{(2\pi)^6}\,
e^{i(\kk^2-\kk'^2)t/2}(k_z+k_z')\kk'\, a(\kk)^*a(\kk')
\int d^2\rhob\,    e^{i(\kappab'-\kappab)\rhob}
      \nonumber\\
&=&  \int \frac{d^3k\, dk_z'}{(2\pi)^4}
e^{i(k_z^2-k_z'^2)t/2}(k_z+k_z')(\kappab+k_z'\ez) a(\kk)^*a(\kappab,k_z')
\,,
\label{e.ver010}
\eea
\labeld{e.ver010}
with $\kappab\equiv k_x\ex+k_y\ey$\@. To verify the significance of $\qdot$, we compute the total change from the beginning of the scattering event until the end. Thus
\bea
\Delta\,\qq &\equiv&\int_{-\infty}^{\infty}\langle\psi|\qdot|\psi\rangle dt
\nonumber\\ &=&
 \int \frac{d^3k\, dk_z'}{(2\pi)^4}
\, a(\kk)^*a(\kappab,k_z')(k_z+k_z')(\kappab+k_z'\ez)
\int_{-\infty}^{\infty}dt\,e^{i(k_z^2-k_z'^2)t/2}
\eea
Now use $\int_{-\infty}^{\infty}e^{-it(u^2-v^2)/2}dt=(2\pi/|u|)\left[\delta(u-v)+\delta(u+v)\right]$\@. Since $(k_z+k_z')$ appears in the integrand, this gives
\bea
\Delta\,\qq &=&
 2\int \frac{d^3k}{(2\pi)^3}
\, a(\kk)^*a(\kk)\kk \sgn(k_z)
\,.
\eea
Interpretation of this result requires a bit of care. Consider $\Delta k_z$\@. Using the symmetry of the wave function ($a(-\kk)=\pm a(\kk)$) it is not difficult to show that the change in $k_z$ is $4\langle k_z\rangle$\@. Slightly more detailed analysis shows that the same is true for the other components of $\kk$\@. Now it should be realized that there is \textit{always} a (no-interaction) scattering. Even if at some early time the particles are moving away from each other, at a yet earlier time they were approaching. Moreover the factor 4 reflects the fact that this is in the center of mass system and that there are two particles each of which has reversed momentum.

\textit{Discussion.~~}
We first comment on three technical issues.

Defining \Mz, the space of the representation, requires the choice of an arbitrary direction. As for other manifolds, limitations may be imposed on the coordinates. For example, the circle, \Sone, has covering space $\Rbb$\@. The representation $\{\theta\,|\,0\le\theta<2\pi\}$ for \Sone\ is unsuitable for defining the velocity of a particle at zero. However, as shown in the calculation of $\Delta \qq$, our choice does not affect the physical result.

The formalism of quantum field theory automatically provides the correct statistics, but seems committed to assigning meaningless names to the indistinguishable particles. Thus the two-particle state could be $a_{\pp_1}^\dagger a_{\pp_2}^\dagger|0\rangle$\@. If one looks at $\psi(\rr_1,\rr_2)\equiv \langle \rr_1\rr_2 |a_{\pp_1}^\dagger a_{\pp_2}^\dagger|0\rangle$, it necessarily has a particular symmetry and lives on $\Rbb^6$. Note though that this does not lead to practical problems: you can calculate energy levels without defining relative momentum.

As for spin, quantum theory introduces a new concept---identity---that complicates a semiclassical limit. No matter how similar two particles are classically, there is no problem defining their relative momentum. This is not the case in quantum theory.

The overall implication of the present work is that the relation between observables and operators does not generally hold. The measurements of momenta discussed at the beginning of this article ($\pp_A$, etc.) were in fact position measurements, done 4 times. Similarly, for time \cite{note:operational}. Surely the relation between operators and observables has been a rich and fruitful perspective in the interpretation of quantum mechanics, but as we have seen, its generality has limitations.


\begin{thebibliography}{15}
\expandafter\ifx\csname natexlab\endcsname\relax\def\natexlab#1{#1}\fi
\expandafter\ifx\csname bibnamefont\endcsname\relax
  \def\bibnamefont#1{#1}\fi
\expandafter\ifx\csname bibfnamefont\endcsname\relax
  \def\bibfnamefont#1{#1}\fi
\expandafter\ifx\csname citenamefont\endcsname\relax
  \def\citenamefont#1{#1}\fi
\expandafter\ifx\csname url\endcsname\relax
  \def\url#1{\texttt{#1}}\fi
\expandafter\ifx\csname urlprefix\endcsname\relax\def\urlprefix{URL }\fi
\providecommand{\bibinfo}[2]{#2}
\providecommand{\eprint}[2][]{\url{#2}}

\bibitem[{\citenamefont{Gottfried}(1966)}]{gottfried}
\bibinfo{author}{\bibfnamefont{K.}~\bibnamefont{Gottfried}},
  \emph{\bibinfo{title}{Quantum Mechanics}} (\bibinfo{publisher}{Benjamin},
  \bibinfo{address}{New York}, \bibinfo{year}{1966}), \bibinfo{edition}{1st}
  ed.

\bibitem[{\citenamefont{Griffiths}(1984)}]{griffiths}
\bibinfo{author}{\bibfnamefont{R.~B.} \bibnamefont{Griffiths}},
  \bibinfo{journal}{J. Stat.\ Phys.} \textbf{\bibinfo{volume}{36}},
  \bibinfo{pages}{219} (\bibinfo{year}{1984}).

\bibitem[{\citenamefont{Muga et~al.}(2002)\citenamefont{Muga, Sala-Mayato, and
  Egusquiza}}]{muga}
\bibinfo{author}{\bibfnamefont{J.~G.} \bibnamefont{Muga}},
  \bibinfo{author}{\bibfnamefont{R.}~\bibnamefont{Sala-Mayato}},
  \bibnamefont{and} \bibinfo{author}{\bibfnamefont{I.~L.}
  \bibnamefont{Egusquiza}}, \emph{\bibinfo{title}{Time in Quantum Mechanics}}
  (\bibinfo{publisher}{Springer-Verlag}, \bibinfo{address}{Berlin},
  \bibinfo{year}{2002}).

\bibitem[{not({\natexlab{a}})}]{note:operational}
\bibinfo{note}{\noteoperational}.

\bibitem[{\citenamefont{Bonneau et~al.}(2001)\citenamefont{Bonneau, Faraut, and
  Valent}}]{bonneau}
\bibinfo{author}{\bibfnamefont{G.}~\bibnamefont{Bonneau}},
  \bibinfo{author}{\bibfnamefont{J.}~\bibnamefont{Faraut}}, \bibnamefont{and}
  \bibinfo{author}{\bibfnamefont{G.}~\bibnamefont{Valent}},
  \bibinfo{journal}{Am. J. Phys.} \textbf{\bibinfo{volume}{69}},
  \bibinfo{pages}{322} (\bibinfo{year}{2001}).

\bibitem[{\citenamefont{Messiah}(1961)}]{messiah}
\bibinfo{author}{\bibfnamefont{A.}~\bibnamefont{Messiah}},
  \emph{\bibinfo{title}{Quantum Mechanics}}, vol.~\bibinfo{volume}{1}
  (\bibinfo{publisher}{North-Holland}, \bibinfo{address}{Amsterdam},
  \bibinfo{year}{1961}), \bibinfo{note}{p.\ 346}.

\bibitem[{not({\natexlab{b}})}]{note:multiple}
\bibinfo{note}{\notemultiple}.

\bibitem[{\citenamefont{Schulman}(1967)}]{thesis}
\bibinfo{author}{\bibfnamefont{L.~S.} \bibnamefont{Schulman}}, Ph.D. thesis,
  \bibinfo{school}{Princeton University} (\bibinfo{year}{1967}).

\bibitem[{\citenamefont{Schulman}(1968)}]{pispin}
\bibinfo{author}{\bibfnamefont{L.~S.} \bibnamefont{Schulman}},
  \bibinfo{journal}{Phys.\ Rev.} \textbf{\bibinfo{volume}{176}},
  \bibinfo{pages}{1558} (\bibinfo{year}{1968}).

\bibitem[{\citenamefont{Schulman}(1971)}]{approximate}
\bibinfo{author}{\bibfnamefont{L.~S.} \bibnamefont{Schulman}},
  \bibinfo{journal}{J. Math.\ Phys.} \textbf{\bibinfo{volume}{12}},
  \bibinfo{pages}{304} (\bibinfo{year}{1971}).

\bibitem[{\citenamefont{Laidlaw and DeWitt-Morette}(1971)}]{laidlaw}
\bibinfo{author}{\bibfnamefont{M.~G.~G.} \bibnamefont{Laidlaw}}
  \bibnamefont{and}
  \bibinfo{author}{\bibfnamefont{C.}~\bibnamefont{DeWitt-Morette}},
  \bibinfo{journal}{Phys.\ Rev.\ D} \textbf{\bibinfo{volume}{3}},
  \bibinfo{pages}{1375} (\bibinfo{year}{1971}).

\bibitem[{\citenamefont{Schulman}(1981)}]{lsspibookallshort}
\bibinfo{author}{\bibfnamefont{L.~S.} \bibnamefont{Schulman}},
  \emph{\bibinfo{title}{Techniques and Applications of Path Integration}}
  (\bibinfo{publisher}{Wiley}, \bibinfo{address}{New York},
  \bibinfo{year}{1981}), \bibinfo{note}{{W}iley Classics 1996; Dover 2005, with
  supplements}.

\bibitem[{\citenamefont{Hegerfeldt et~al.}(2006)\citenamefont{Hegerfeldt,
  Neumann, and Schulman}}]{detectormodelmovingparticles}
\bibinfo{author}{\bibfnamefont{G.~C.} \bibnamefont{Hegerfeldt}},
  \bibinfo{author}{\bibfnamefont{J.~T.} \bibnamefont{Neumann}},
  \bibnamefont{and} \bibinfo{author}{\bibfnamefont{L.~S.}
  \bibnamefont{Schulman}}, \bibinfo{journal}{J. Phys.\ A}
  \textbf{\bibinfo{volume}{39}}, \bibinfo{pages}{14447} (\bibinfo{year}{2006}).

\bibitem[{\citenamefont{Peres}(1980)}]{peres}
\bibinfo{author}{\bibfnamefont{A.}~\bibnamefont{Peres}}, \bibinfo{journal}{Am.
  J. Phys.} \textbf{\bibinfo{volume}{48}}, \bibinfo{pages}{552}
  (\bibinfo{year}{1980}).

\bibitem[{\citenamefont{Salecker and Wigner}(1956)}]{wigner}
\bibinfo{author}{\bibfnamefont{H.}~\bibnamefont{Salecker}} \bibnamefont{and}
  \bibinfo{author}{\bibfnamefont{E.~P.} \bibnamefont{Wigner}},
  \bibinfo{journal}{Phys. Rev.} \textbf{\bibinfo{volume}{109}},
  \bibinfo{pages}{571} (\bibinfo{year}{1956}).

\end{thebibliography}
\end{document}